\documentstyle[12pt]{article}
\title{\bf 1+1 string with quarks at the ends revisited}
\author{Yu.S.Kalashnikova\thanks{e-mail: yulia@vxitep.itep.ru},
A.V.Nefediev\thanks{e-mail: nefediev@vxitep.itep.ru}}
\date{\it Institute of Theoretical and Experimental Physics, 117259, Moscow,
Russia}
\newcommand{\xp}{x_{\perp}}
\newcommand{\ds}{\displaystyle}
\newcommand{\be}{\begin{equation}}
\newcommand{\ee}{\end{equation}}
\begin{document}
\maketitle

\begin{abstract}
The canonical description is presented for the string with pointlike masses
at the ends in 1+1 dimensions in two different gauges: in the proper time
gauge and in the light cone one. The classical canonical transformation
is written out explicitly, which relates physical variables in both gauges,
and equivalence of two classical theories is demonstrated in such a way.
Both theories are quantized, and it is shown that quantum theories are not
unitary equivalent. It happens due to the fact that the canonical
transformation depends on interaction. The quantum Poincar{\' e} algebra
proves to be closed in both cases, so that the requirement of Poincar{\'e}
covariance is not able to distinguish between two versions of the theory.
\end{abstract}

{\parindent=0cm {\bf 1.}\hspace*{0.4cm}The
modern quark constituent model views meson as a
$q\bar q$ pair connected with a string in its ground state, and
vibrational string excitations are responsible for formation of hybrid mesons.}
Such a picture is qualitatively consistent with the expectations based on
QCD at large distances, but it is by no means straightforward to formulate
a model in a more or less rigorous way. An open string {\it per se} is not a
simple dynamical object, and putting quarks at the ends of it makes the
problem extremely complicated even at the classical level.

Many years ago a much more simple model of a string with quarks in 1+1
dimensions was considered \cite{1, 2}. As there is no room for transverse
vibrations in 1+1, the problem is solvable exactly, and classical solutions
were presented in timelike and lightlike gauges. Both solutions were
quantized, and as the result the Bethe--Salpiter equation for the timelike
version and the 't~Hooft equation for the lightlike one emerged. The quantum
spectra of these equations are different, though the quasiclassical limits
coincide. Therefore the question arises of which theory is the correct one.
Natural way to resolve this question is to check if the quantum Poincar{\'e}
algebra is closed in each case. It is easy to verify that it is so for the
lightlike gauge fixing, while for the timelike gauge it is not so at
first glance. In a recent paper \cite{3}, where the similar situation of
1+1 QED was considered, this phenomenon was described as an example of
quantum Poincar{\'e} anomaly. However, it was argued in \cite{2} involving
rather general considerations that the algebra should close in the timelike
gauge if proper canonical variables are chosen. The explicit form of
these variables was not given in \cite{2}, and, as far as we know, has not
been found up to now.

In the present letter we consider the 1+1 string with quarks in the proper
time gauge which was advocated a lot by Rohrlich (see \cite{4} and
references therein). In contrast to laboratory timelike gauge it allows to
separate the centre-of-mass motion in a simple way, and to establish the
properly commuting variables. The equations in the proper time gauge
coincide with the ones in the timelike one in the rest frame, while
the difficulties encountered in the laboratory timelike gauge in an
arbitrary frame are avoided. We present the classical canonical
transformation relating the coordinate and momentum in the proper time
and lightlike gauges, and speculate on how it happens that two classically
equivalent theories are not unitary equivalent at the quantum level.

We demonstrate that quantum Poincar{\'e} algebra is closed in the proper
time gauge, and therefore we confirm the observation of \cite{2} that
there is no quantum anomaly, and quantum Poincar{\'e} covariance is not
able to single out the "true" theory. In conclusion, a possible way out
of such a situation is discussed.
\medskip

{\parindent=0cm {\bf 2.}\hspace*{0.4cm}The action of a string with quarks
at the ends is}
\be
S=\int_{\tau_1}^{\tau_2}d\tau L(\tau ),\;
L(\tau)=-m_1\sqrt{\dot{x}_1^2}-m_2\sqrt{\dot{x}_2^2}
-\sigma\int_0^1d\beta\sqrt{(\dot{w}w')^2-\dot{w}^2w'^2},
\label{1}
\ee
where $w_{\mu}=w_{\mu}(\tau,\beta)\;(\mu=0,\;1)$ are the coordinates of
the string world surface, $\dot{w}_{\mu}=\frac{\partial w_{\mu}}
{\partial\tau}$,
$w'_{\mu}=\frac{\partial w_{\mu}}
{\partial\beta}$, and
$x_{1\mu}(\tau)=w_{\mu}(\tau,\;0)$,
$x_{2\mu}(\tau)=w_{\mu}(\tau,\;1)$.
Action (\ref{1}) is invariant under the usual group of $\tau$ and $\beta$
reparametrizations
$$
\begin{array}{lllll}
\tau&\to\tilde\tau(\tau,\beta),&\tilde\tau(\tau_1,\beta)=\tau_1&
\tilde\tau(\tau_2,\beta)=\tau_2\\
%{}\\
\beta&\to\tilde\beta(\tau,\beta),&\tilde\beta(\tau,0)=0&
\tilde\beta(\tau,1)=1
\end{array}
$$
We find it more convenient to rewrite the Lagrangian (\ref{1}) introducing
the so-called einbein fields (see {\it e.g.} \cite{5}):
\be
L(\tau)=-\frac{m_1^2}{2\mu_1}-\frac{\mu_1\dot{x}_1^2}{2}
-\frac{m_2^2}{2\mu_2}-\frac{\mu_2\dot{x}_2^2}{2}
-\sigma\int_0^1d\beta\sqrt{(\dot{w}w')^2-\dot{w}^2w'^2},
\label{2}
\ee

Theories (\ref{1}) and (\ref{2}) are equivalent, while the einbein fields
formalism is rather helpful in solving the problem of the centre-of-mass
motion separation~\cite{6}.

First we, following \cite{1}, note that if the evolution parameter in
(\ref{1}) is specified by fixing the gauge in $\tau$--reparametrization
group as
\be
(nw)=\tau,
\label{3}
\ee
where timelike $(n^2=1)$ or lightlike $(n^2=0)$ two-vector does not depend on
$\tau$, then one can fix the gauge in $\beta$--reparametrization group by
choosing the uniform gauge:
\be
\frac{\partial^2 w_{\perp}}{\partial \beta^2}=0,\quad w_{\perp}=
(1-\beta)x_{1\perp}+\beta x_{2\perp},
\label{4}
\ee
where sign $\perp$ indicates the components of the corresponding vectors
transverse with respect to $n_{\mu}$.

Equations (\ref{3}) and (\ref{4}) supplied by some specification of the
vector $n_{\mu}$ fix the gauges completely, and, in accordance with general
Dirac considerations \cite{7}, no first class constraints appear in the
Hamiltonian formulation of the problem. However, as the einbein fields are
introduced, the second class constraints appear, which have different
properties for different choices of~$n_{\mu}$.
\smallskip

{\parindent=0cm {\bf 3.}\hspace*{0.4cm}{\it Timelike gauge}.}
Lagrangian (\ref{2}) in the timelike gauge takes the form
\be
L(\tau)=-\frac{m_1^2}{2\mu_1}-\frac{\mu_1}{2}+\frac{\mu_1\dot{x}_{1\perp}^2}{2}
-\frac{m_2^2}{2\mu_2}-\frac{\mu_2}{2}+\frac{\mu_2\dot{x}_{2\perp}^2}{2}
-\sigma|x_{1\perp}-x_{2\perp}|.
\label{5}
\ee

To separate the centre-of-mass motion we introduce new variables
\be
\begin{array}{ll}
x_{\perp}=x_{1\perp}-x_{2\perp},&
\quad X_{\perp}=\frac{\ds\mu_i}{\ds\mu_1+\mu_2}x_{1\perp}+\frac{\ds\mu_2}
{\ds\mu_1+\mu_2}x_{2\perp}\\
{}\\
\zeta=\frac{\ds\mu_1}{\ds\mu_1+\mu_2},&\quad M=\mu_1+\mu_2
\end{array}
\label{6}
\ee

In terms of these new variables the Lagrangian becomes
\be
L=-\frac{m_1^2}{2M\zeta}-\frac{m_2^2}{2M(1-\zeta)}-\frac{M}{2}+
\frac12M(\dot{X}_{\perp}-\dot{\zeta}x_{\perp})^2+\frac12M\zeta(1-\zeta)
\dot{x}^2_{\perp}-\sigma|x_{\perp}|
\ee

The canonical momenta defined as
\be
P_{\perp}=\frac{\partial L}{\partial \dot{X}_{\perp}},\quad
p_{\perp}=\frac{\partial L}{\partial \dot{x}_{\perp}},\quad
\Pi=\frac{\partial L}{\partial \dot{M}},\quad
\pi=\frac{\partial L}{\partial \dot{\zeta}}
\ee
give the Hamilton function
\be
H=H_0+\Lambda\varphi_1+\lambda\varphi_2,
\ee
$$
H_0=\frac{\varepsilon_1^2}{2M\zeta}+\frac{\varepsilon_2^2}{2M(1-\zeta)}+
\frac{P_{\perp}^2}{2M}+\frac{M}{2}+\sigma|x_{\perp}|,\quad
\varepsilon_i^2=m_i^2+p^2_{\perp}
$$
and primary constraints
\be
\varphi_1=\Pi,\quad\varphi_2=\pi+P_{\perp}x_{\perp},
\ee
which are added to the Hamiltonian $H_0$ with Lagrange multipliers $\Lambda$
and $\lambda$. Secondary constraints are
\be
\begin{array}{l}
\varphi_3=\{\varphi_1H\}=-
\frac{\ds\varepsilon_1^2}{\ds 2M^2\zeta}-\frac{\ds\varepsilon_2^2}
{\ds2M^2(1-\zeta)}-
\frac{\ds P_{\perp}^2}{\ds 2M^2}+\frac{\ds 1}{\ds 2}\\
{}\\
\varphi_4=\{\varphi_2H\}=-
\frac{\ds\varepsilon_1^2}{\ds 2M\zeta^2}+\frac{\ds\varepsilon_2^2}
{\ds2M(1-\zeta)^2}-
\frac{\ds p_{\perp}P_{\perp}}{\ds M\zeta(1-\zeta)}.
\end{array}
\ee

No further constraints appear, as the requirements
$\{\varphi_3H\}=0$ and $\{\varphi_4H\}=0$ are actually the equations defining
the Lagrange multipliers $\Lambda$ and $\lambda$. It is easy to check
that the set of constraints $\varphi_i,\;i=1\ldots 4$ is of the second
class, and one can use it to exclude the redundant variables and
to define the Dirac brackets for physical variables. Choosing as physical
variables the coordinates and momenta $X_{\perp},\;x_{\perp}\;P_{\perp}$
and $p_{\perp}$ we arrive at the physical Hamiltonian
\be
H_{\rm phys}=\sqrt{P_{\perp}^2+\varepsilon_1^2+\varepsilon_2^2
+2\sqrt{\varepsilon_1^2\varepsilon_2^2+p_{\perp}^2P_{\perp}^2}}
+\sigma|x_{\perp}|.
\ee

Not only the total momentum $P_{\perp}$ enters the Hamiltonian in a very
unpleasant way, but also the Dirac brackets for the physical variables are
distorted due to the presence of the second class constraints:
\be
\begin{array}{l}
\{X_{\perp}p_{\perp}\}_D=\frac{\ds p_{\perp}P_{\perp}}{\ds M^2},\;\;
\{X_{\perp}x_{\perp}\}_D=\frac{\ds x_{\perp}P_{\perp}}{\ds M^2}+
\frac{\ds x_{\perp}p_{\perp}}{\ds M^2}\left(\frac{\ds 1-\zeta}
{\ds \zeta}-\frac{\ds \zeta}{\ds 1-\zeta}\right),\\
{}\\
\{p_{\perp}x_{\perp}\}_D=1-\frac{\ds P^2_{\perp}}{\ds M^2}-
\frac{\ds p_{\perp}P_{\perp}}{\ds M^2}\left(\frac{\ds 1-\zeta}
{\ds \zeta}-\frac{\ds \zeta}{\ds 1-\zeta}\right).
\end{array}
\label{13}
\ee

The noncanonical form of brackets (\ref{13}) is not an artifact of the
einbein field formalism.
Such a situation is well--known in the relativistic quantum mechanics
\cite{8}. To bring brackets (\ref{13}) into the canonical form the
centre-of-mass Newton--Wigner variable $Q_{\perp}$ should be defined,
which properly commutes with the Hamiltonian:
\be
\dot{Q_{\perp}}=\{HQ_{\perp}\}=\frac{P_{\perp}}{H}.
\ee

Search for such variables and corresponding new internal ones is not a
trivial problem even for the case of two free particles \cite{2, 6}.
As it was already mentioned, this problem for the case of two
particles interacting via string has not been solved up to now.

There are two reasonable choices of the timelike vector $n_{\mu}$ in
(\ref{3}). One way is to identify the evolution parameter $\tau$ with
the laboratory time, setting $w_0=\tau$. With such a gauge choice we
are faced with all the difficulties described above. A roundabout way is
to identify the parameter $\tau$ with the proper time, taking
$n_{\mu}=\frac{P_{\mu}}{\sqrt{P^2}}$. As in this gauge $P_{\perp}=0$,
the dynamics of the center-of-mass motion disappears from the Hamiltonian
and constraints in a trivial way as it should be in the co-moving frame,
so the physical Hamiltonian becomes
\be
H=\varepsilon_1+\varepsilon_2+\sigma |x_{\perp}|,
\label{15}
\ee
with $\{p_{\perp}x_{\perp}\}=1$.

Now it is possible to perform the transformation into laboratory timelike
gauge in {\it the centre-of-mass variables only}, leaving the internal ones
untouched. The Hamiltonian in this gauge is
\be
P_0=\sqrt{P_1^2+H^2},
\ee
where $H$ is taken from equation (\ref{15}) with
\be
p_{\perp}=-(p_{\mu}e_{\mu}^{(1)}),\quad
x_{\perp}=-(x_{\mu}e_{\mu}^{(1)}),
\label{17}
\ee
and the diade of vectors $e_{\mu}^{(\alpha)}$, $\alpha=0,\;1$ is defined as
\be
e_{\mu}^{(0)}=\frac{P_{\mu}}{\sqrt{P^2}},\quad
(e_{\mu}^{(0)}e_{\mu}^{(1)})=0,\quad
(e_{\mu}^{(1)}e_{\mu}^{(1)})=-1.
\label{18}
\ee

In spite of the fact that the internal variables (\ref{17}) depend on the
total momentum, the brackets have the canonical form
\be
\begin{array}{l}
\{P_1X_1\}=\{p_{\perp}x_{\perp}\}=1,\\
{}\\
\{P_1x_{\perp}\}=\{P_1p_{\perp}\}=\{X_1x_{\perp}\}=\{X_1p_{\perp}\}=0,
\end{array}
\label{19}
\ee
that can be easily verified using the explicit form of the diade vectors
(\ref{18}).

The Poincar{\'e} group generators are written as
\be
P_0=\sqrt{P_1^2+H^2},\quad P_1=P_1,\quad M_{01}=-\tau P_1+X_1P_0,
\ee
and commute as
\be
\{P_0P_1\}=0,\quad\{P_0M_{01}\}=P_1,\quad\{P_1M_{01}\}=P_0
\label{21}
\ee

The theory can be canonically quantized in terms of variables
$P_1,\;X_1,\;p_{\perp}$ and $x_{\perp}$ by the correspondence principle
$\{AB\}\to i[\hat{A}\hat{B}]$, where $[\hat{A}\hat{B}]$ is the quantum
commutator. The Poincar{\'e} algebra (\ref{21}) continues to hold at the
quantum level, if one just hermitizes the last term in $M_{01}$ as
$X_1P_0\to\frac12(\hat{X}_1\hat{P}_0+\hat{P}_0\hat{X}_1)$.
\medskip

{\parindent=0cm {\bf 4.}\hspace*{0.4cm}{\it Lightlike gauge.}} The
lightlike gauge
is organized in a drastically different way in the einbein field
formalizm. Choosing vector $n_{\mu}$ in (\ref{3})  by setting
$\tau=\frac{1}{\sqrt{2}}(w_1-w_0)=w_+$, we arrive at the Lagrangian
\be
L=-\frac{m_1^2}{2\mu_1}+\mu_1\dot{x}_{1-}-
\frac{m_2^2}{2\mu_2}+\mu_2\dot{x}_{2-}
-\sigma|x_{1-}-x_{2-}|.
\label{22}
\ee

Performing the transformation of coordinates similar to (\ref{6})
\be
\begin{array}{ll}
x_{-}=x_{1-}-x_{2-},&
\quad X_{-}=\frac{\ds\mu_i}{\ds\mu_1+\mu_2}x_{1-}+\frac{\ds\mu_2}
{\ds\mu_1+\mu_2}x_{2-}\\
{}\\
y=\frac{\ds\mu_1}{\ds\mu_1+\mu_2},&\quad M=\mu_1+\mu_2
\end{array}
\label{23}
\ee
we rewrite the Lagrangian as
\be
L=-\frac{m_1^2}{2yM}-\frac{m_2^2}{2(1-y)M}+M(\dot{X}_{-}-\dot{y}x_{-})
-\sigma|x_{-}|
\ee
and the Hamiltonian and primary constraints take the form
\be
H=\frac{m_1^2}{2yM}+\frac{m_2^2}{2(1-y)M}+\sigma|x_{-}|+\Lambda\varphi_1+
\lambda\varphi_2+E\varphi_3+e\varphi_4
\ee
\be
\varphi_1=\Pi,\quad\varphi_2=\pi+Mx_-,\quad\varphi_3=P_+-M,\quad\varphi_4=p_+,
\ee
where $P_+,\;p_+,\;\Pi$ and $\pi$ are the momenta conjugated to the
coordinates $X_-,\;x_-,\;M$ and $y$ correspondingly.

In contrast to the timelike case, no secondary constraints appear, as the
equations $\{\varphi_iH\}=0$ define the Lagrange multipliers $\Lambda,\;
\lambda,\;E$ and $e$. Moreover, the Dirac brackets for the physical variables
which are chosen to be $P_+,\;X_-,\;y$ and $\pi$ coincide with the Poisson
ones and therefore have the canonical form
\be
\begin{array}{l}
\{P_+X_-\}=\{\pi y\}=1,\\
{}\\
\{P_+\pi\}=\{P_+y\}=\{X_1\pi\}=\{X_1y\}=0.
\end{array}
\label{27}
\ee

The physical Hamiltonian takes the form
\be
H=\frac{1}{2P_+}\left(\frac{m_1^2}{y}+\frac{m_2^2}{1-y}\right)+
\sigma\frac{|\pi|}{P_+}.
\ee

The quantization is straightforward with brackets (\ref{27}), and the
quantum spectrum is given by the mass squared operator
\be
{\cal M}^2=\frac{m_1^2}{y}+\frac{m_2^2}{1-y}+2\sigma |\pi|,
\ee
coinciding with the one derived by 't Hooft for the 1+1 QCD in the large
$N_c$ limit \cite{9}.

The Poincar{\'e} generators are given by
\be
P_-=\frac{1}{2P_+}\left(\frac{m_1^2}{y}+\frac{m_2^2}{1-y}\right)+
\sigma\frac{|\pi|}{P_+},\; P_+=P_+,\;
M_{+-}=-\tau P_-+\frac12(X_-P_++P_+X_-),
\ee
and properly commute as
\be
\{P_+P_-\}=0,\quad\{P_-M_{+-}\}=-P_-,\quad\{P_+M_{+-}\}=P_+
\label{31}
\ee
both at classical and quantum levels.
\medskip

{\parindent=0cm {\bf 5.}\hspace*{0.4cm}The fact that theory (\ref{1}) in the
lightlike gauge is consistent with the requirement of quantum
Poincar{\'e} invariance was established in \cite{2}}. The quantum Poincar{\'e}
invariance is apparently lost if one sticks with the single particle
coordinates and momenta $x_1,\;x_2\;p_1$ and $p_2$ in the laboratory
timelike gauge \cite{2}. The same situation takes place in 1+1 QED \cite{3}.
Our findings are by no means in contradiction with these statements: as it
was anticipated in \cite{2} the set of canonical variables should exist which
differs from the single particle ones and for which quantum algebra is closed
as well as the classical one, and we have found such a set.

It was argued in \cite{2} that as we are dealing with the same theory
(\ref{1}) in differeent gauges, then a canonical transformation should
exist which relates the variables in both gauges. Here we
give the generating function for this transformation for a more simple case
of equal masses, $m_1=m_1=m$:
$$
F(\xp ,y)=\frac{(1-2y)\xp}{16y(1-y)}\left(\sigma |\xp|+
\sqrt{(1-2y)^2\sigma^2\xp^2+16m^2y(1-y)}\right)
$$
\be
\hspace*{1cm}+\frac{m^2}{\sigma}sign\left(\frac12-y\right)ln\frac{
\sigma \xp|1-2y|+\sqrt{(1-2y)^2\sigma^2\xp^2+16m^2y(1-y)}
}{4m\sqrt{y(1-y)}}
\label{32}
\ee
with $\pi=\frac{\partial F}{\partial y},\;p_{\perp}=
\frac{\partial F}{\partial x_{\perp}}$.
In spite of the frightening appearance function (\ref{32}) possesses
good analytical properties at points $\xp=0$ and $y=\frac12$.
The generating function for the case of $m_1\ne m_2$ is much more complicated
and we don't write it down here. Nevertheless, in terms of \lq\lq old"
variables
$p_{\perp}$ and $x_{\perp}$ \lq\lq new" ones $\pi$ and $y$ are expressed as
\be
y=\frac{{\cal M}^2-m_2^2+m_1^2}{2{\cal M}^2}-\frac{p_{\perp}}{\cal M}
\label{325}
\ee
\be
\pi=\frac{x_{\perp}}{2\sigma|\xp|}\left[
{\cal M}^2-\frac{m_1^2}{y(p_{\perp},\xp)}-\frac{m_2^2}{1-y(p_{\perp},\xp)}
\right]
\label{33}
\ee

$$
{\cal M}=\varepsilon_1+\varepsilon_2+\sigma |x_{\perp}|,
$$

It follows from equations (\ref{325}), (\ref{33}) that
\be
{\cal M}^2=(\varepsilon_1+\varepsilon_2+\sigma |x_{\perp}|)^2=
\frac{m_1^2}{y}+\frac{m_2^2}{1-y}+2\sigma |\pi|.
\ee

Note that transformation (\ref{325}), (\ref{33}) is not singular
at $\xp=0$, so that
\be
\{\pi,y\}_{p_{\perp},\xp}=\frac{\partial \pi}{\partial p_{\perp}}
\frac{\partial y}{\partial x_{\perp}}-
\frac{\partial \pi}{\partial x_{\perp}}
\frac{\partial y}{\partial p_{\perp}}=1
\ee
everywhere including $\xp=0$, {\it i.e.} this transformation is canonical.

The existence of canonical transformation (\ref{325}), (\ref{33})
proves the classical equivalence of the theories. But what about the quantum
equivalence? The quasiclassical spectra of the Bethe--Salpiter and the
't Hooft equations coincide, as it was demonstrated by calculating the
corresponding
Bohr--Zommerfeld integrals, yielding the celebrated Regge-like behaviour in
the $m_1,\;m_2\to 0$ limit,
\be
{\cal M}^2=2\pi\sigma n,
\ee
but the quantum spectra are not the same, as it could be easily seen either
by inspection of numerical solutions or by performing nonrelativistic limit
and comparing the first relativistic corrections. The transformation
(\ref{325}), (\ref{33}) is neither contact, but mixes coordinates and momenta,
nor it is linear.
Moreover, it essentially depends on interaction.
So there should not be quantum unitary equivalence of the
theories.

The physical reasons are also quite clear: quantum mechanics does not deal
with backward motion of particles, and negative energy states are removed
by truncating the Fock space. It was shown in \cite{3} that the boost operator
commutes with the projection operator on the positive energy states in the
lightlike gauge. But the proper time gauge possesses the same property by
construction: there is no particle--antiparticle production in the
co-moving frame. Thus the theories are quantized in differently truncated
Fock spaces, and no surprise that the spectra do not coincide. The
embarrassing question is what version is the correct one.

The Bethe--Salpiter equation has a long history of successful
phenomenological
applications, while the 't Hooft equation has a respectable field theory
background behind it. The original derivation \cite{9} was performed by
summing up planar graphs and employing the light-cone gauge condition
$A_-=0$. It was followed by speculations \cite{10} that the result might be
an artifact of the lightcone gauge, and might not be reproducable in other
axial gauges. Then the problem was resolved by deriving the 't Hooft equation
employing the Coulomb gauge \cite{11}. On the other hand, action (\ref{1})
arises in a natural way in QCD in the Wilson loop approach to $q\bar q$ Green
function in neglection of quark loops \cite{12}, so the
string spectrum in 1+1 dimensions
should coincide with the one of the 1+1 QCD.

Of course, the 1+1 QCD has much more content than simple quantum mechanical
reduction (\ref{1}). Still one issue could be relevant to the puzzle:
QCD deals with fermionic quarks, while the quarks in (\ref{1}) are spinless.
The two--dimensional quantum mechanical models are poor in symmetries, and
it might be helpful to place spinning quarks at the ends of the string,
addind in such a way extra degrees of freedom and, consequently, extra
symmetries which should hold at the quantum level \cite{13}. We hope that
the latter requirement would allow to single out the \lq\lq true" theory.
\medskip

We acknowledge useful discussions with A.A.Abrikosov Jr., K.G.Boreskov,
O.V.Kan\-che\-li, L.A.Kondratyuk and A.V.Smilga.

The work is supported by grant $N^{\underline{0}}$ 96-02-19184a of
Russian Fundamental Research Foundation and by INTAS 94-2851.


\begin{thebibliography}{99}
\bibitem{1}I.Bars, A.J.Hanson, Phys.Rev. {\bf D13} (1976), 1744\\
W.A.Bardeen, I.Bars, A.J.Hanson, R.D.Peccei, Phys.Rev. {\bf D13} (1976), 2364
\bibitem{2}W.A.Bardeen, I.Bars, A.J.Hanson, R.D.Peccei, Phys.Rev.
{\bf D14} (1976), 2193
\bibitem{3}S.Lenz, B.Schreiber, Phys.Rev. {\bf D53} (1996), 960
\bibitem{4}F.Rohrlich, Ann.Phys. {\bf 117} (1979), 292\\
M.J.King, F.Rohrlich, Ann.Phys. {\bf 130} (1980), 350
\bibitem{5}L.Brink, P.Di Vecchia, P.Howe, Nucl.Phys. {\bf B118} (1977), 76
\bibitem{6}Yu.S.Kalashnikova, A.V.Nefediev, ITEP-45-96, hep-ph/9611361,
Journ.Atom.Nucl., in press
\bibitem{7}P.A.M.Dirac, \lq\lq Lectures on Quantum mechanics", Belter Graduate
School of Science, Yeshiva University, New York (1964)
\bibitem{8}M.H.L.Pryce, Proc,Roy.Soc, London, Ser.{\bf A150} (1935), 166\\
A.J.Hanson, T.Regge, Ann.Phys. {\bf 87} (1974), 498
\bibitem{9}G.'t Hooft, Nucl.Phys. {\bf B75} (1974), 461
\bibitem{10}Y.Frishman, C.T.Sachrajda, H.D.I.Abarbanel, R.Blankenbecler,
Phys.Rev. {\bf D15} (1977), 2275
\bibitem{11}I.Bars, M.B.Green, Phys.Rev. {\bf D17} (1978), 537
\bibitem{12}A.Yu.Dubin, A.B.Kaidalov, Yu.A.Simonov, Phys.Lett. {\bf B323}
(1994), 41, {\bf B343} (1995), 310
\bibitem{13}A.V.Smilga, Nucl.Phys. {\bf B292} (1987), 363\\
A.V.Smilga, private communication

\end{thebibliography}
\end{document}